\documentclass{ws-procs9x6}
\begin{document}

\title{Coherent State Measures and the Extended Dobi\'{n}ski relations}
\author{Karol A. Penson and Allan I. Solomon}
\address{Laboratoire de Physique Théorique des Liquides,\\Université Paris VI, 75252 Paris Cedex 05,
  France\\{E-mail:penson@lptl.jussieu.fr}\\{E-mail:a.i.solomon@open.ac.uk}
}
\maketitle
\abstracts{
Conventional Bell and Stirling numbers arise naturally in the normal ordering of simple monomials in boson operators.  By extending this process we obtain generalizations of these combinatorial numbers, defined as coherent state matrix elements of arbitrary  monomials, as well as the associated Dobi\'{n}ski relations.  These Bell-type numbers may be  considered as power moments and give rise to positive measures which allow the explicit construction of new classes of coherent states.
}

\section{Introduction}
A defining characteristic of coherent states is the {\em resolution of unity} property, which is another expression of the classical Stieltjes moment problem. For the conventional coherent states of quantum optics the moments are simply $n!$, in some sense the simplest combinatorial numbers, leading to a simple exponential weight function. Not many other solutions to the moment problem are known, and  it is in a sense not surprising that one should seek for solutions in the realm of other combinatorial integers. One such set is the Bell and Stirling numbers, which we define explicitly later, and  which arise naturally in the normal ordering properties of boson operators.   

The standard boson commutation relation 
$
[a,a^{\dagger}]=1
$
 can be {\em formally} realised by identifying $a= \frac{d}{dx}$ and $a^{\dagger}=x$, since
 $ 
[\frac{d}{dx},x] = 1.
$
In the present note we shall use both of the above forms.  Integer sequences arise naturally  when considering the action of 
$(x \, {d}/{dx})^{n}$ on $f(x)$ as the following low order examples reveal:
\begin{eqnarray}
(x \frac{d}{dx}) f(x) & = & x f'(x)  \\ 
(x \frac{d}{dx})^{2}f(x) & = & x f'(x) +x^{2} f''(x)  \\ 
(x \frac{d}{dx})^{3}f(x) & = & x f'(x) +3 x^{2} f''(x) + x^{3} f'''(x) \; \; \; \rm{etc.,}
 \label{dd} \end{eqnarray}
which in general can be written as 
\begin{equation} 
(x\,\frac{d}{dx})^{n}f(x) =\sum_{k=1}^{n} S(n,k) x^{k} ({d}/{dx})^{k}f(x)
\label{s1}
\end{equation}
or, alternatively\cite{kat1}
\begin{equation} 
(a^{\dagger}a)^{n} =\sum_{k=1}^{n} S(n,k) (a^{\dagger})^{k} a^{k}.
\label{s2}
\end{equation}
The {\em Stirling numbers of the second kind} $S(n,k)$ appearing in Eqs.(\ref{dd}) and (\ref{s2}) have been known for over 250 years\cite{yab}.  Eq.(\ref{s2}) exemplifies the {\em normal ordering problem}, that is, finding the form of $(a^{\dagger}a)^{n}$ with the powers of $a$ on the right. Although explicit expressions for $S(n,k)$ are known\cite{com}, of particular interest here are the {\em Bell numbers} $B(n)$ given by the sums
\begin{equation} 
B(n) =\sum_{k=1}^{n} S(n,k), \; \; \; \; \; \; \; n=1,2,\ldots
\label{b1}
\end{equation}
with $B(0)=1$ by convention $(S(n,0)=\delta_{n,0})$. A closed-form expression for $B(n)$ can be found by considering the action of $(x {d}/{dx})^{n}$ on a function $f(x)$ having a Taylor expansion around $x$=$0$, i.e. $f(x)=\sum_{k=0}^{\infty}c_{k} x^{k}.$

Applying Eq.(\ref{s1}) to $f(x)$ gives:
\begin{equation}
 (x\frac{d}{dx})^{n}f(x) = \sum_{k=0}^{\infty} c_k k^{n} x^{k}.  \\ 
 \label{ddd} \end{equation}
Now specify  $f(x)=e^{x}$ in Eq.(\ref{ddd}) so that $c_k = 1/k!\; \;$ and deduce from Eq.(\ref{s1}) that
\begin{equation} 
(1/e^{x}) \sum_{k=0}^{\infty} {\frac{k^n}{k!}}x^{k} =\sum_{k=1}^{n} S(n,k)x^{k},
\label{b2}
\end{equation}
which for $x=1$ reduces to
\begin{equation} 
(1/e) \sum_{k=0}^{\infty} \frac{k^n}{k!} =\sum_{k=1}^{n} S(n,k)=B(n).
\label{b3}
\end{equation}
Equations (\ref{b2}) and (\ref{b3}) are the celebrated Dobi\'{n}ski formulas\cite{yab,com,con,wil} which have been the subject of much combinatorial interest. For completeness, we recall  the combinatorial definitions of $B(n)$ and $S(n,k)$: $B(n)$ counts the number of partitions of a set of $n$ distinguishable elements; $S(n,k)$ counts the number of partitions of a set of $n$ distinguishable elements into $k$ {\em non-empty} sets.  Eq.(\ref{b3}) represents the integer $B(n)$ as an infinite series, which is however {\em not} a power series in $n$.  An immediate consequence of Eqs.(\ref{b2}) and (\ref{b3}) is that $B(n)$ is the $n$-th moment of a (singular) probability distribution, consisting of weighted Dirac delta functions located at the positive integers (the so-called Dirac comb) :
\begin{equation} 
B(n) = \int_{0}^{\infty} x^{n} W(x) dx, \; \; \; n=0,1,\ldots
\label{d1}
\end{equation}
where
\begin{equation} 
W(x) =(1/e) \sum_{k=1}^{\infty} {\frac{\delta(x-k)}{k!}}.
\label{d2}
\end{equation}
The discrete measure $W(x)$ serves as a weight function for a family of orthogonal polynomials ${C}_n^{(1)}(x)$, the Charlier polynomials\cite{koe}. The exponential generating function (EGF) of the sequence $B(n)$ can be obtained from Eq.(\ref{b3}) as
\begin{equation} 
e^{(e^{\lambda} -1)} =\sum_{k=0}^{\infty} {B(n) \frac{{\lambda}^n}{n!}}.
\label{EGF}
\end{equation}
This equation  is related via Eq.(\ref{s2}) to a formula giving  the normal ordered form\cite{kla,lou} of $e^{\lambda a^{\dagger} a}$ 
\begin{equation}
e^{\lambda a^{\dagger} a} = {\mathcal N}(e^{\lambda a^{\dagger} a}) = :e^{a^{\dagger} a(e^{\lambda}-1)}: 
\label{norm}
\end{equation} 
 The symbol ${\mathcal N}$ denotes normal ordering and $:A(a^{\dagger},a):$ means expand $A$ in a Taylor series and normally order {\em without} taking account of the commutation relation $[a,a^{\dagger}] = 1$.  We stress that in the derivation\cite{kla,lou} 
of  Eq.(\ref{norm}) no use has been made of the Stirling and Bell numbers. It may readily be seen that Eq.(\ref{EGF}) is the expectation value of Eq.(\ref{norm}) in the coherent state $|z\rangle$ defined by $a|z\rangle=z|z\rangle$ at the value $|z|=1$.
This circumstance has been used recently to re-establish the link between the matrix element $\langle z|e^{\lambda a^{\dagger} a}|z\rangle$ and the properties of Stirling and Bell 
numbers\cite{kat}.
\section{Extending Dobi\'{n}ski formulas}
The purpose of this note is to show that the above results on functions of $a^{\dagger}a$ may be extended to functions of $(a^{\dagger})^{r} a^{s}, \;  (r,s = 1,2,\ldots) $ with $r\geq s$. 

Specifically, we pose the following questions:
\begin{enumerate}
\item What extensions of the conventional Stirling and Bell numbers  occur in the normal ordering of $[(a^{\dagger})^{r} a^{s}]^{n}\;$?
\item Can the generalised Bell numbers $B_{r,s}(n)$ so defined be represented by an infinite series of the type of Eq.(\ref{b3}) - that is, do they satisfy a generalised Dobi\'{n}ski formula ?
\item May one consider the  $B_{r,s}(n)$ as the $n$-th moments of a positive weight function $W_{r,s}(x)$ on the positive half-axis, and may this latter be explicitly obtained ?
\item Can one attach a combinatorial significance to the sequences $\{B_{r,s}(n)\}\;$?
\end{enumerate}
In this note we indicate affirmative answers to the first three questions and a partial answer to the fourth.

To this end we generalize  Eq.(\ref{s2}) by defining for $r \geq s$:
\begin{equation}
[(a^{\dagger})^{r} a^{s}]^{n}=(a^{\dagger})^{n(r-s)}\sum_{k=s}^{ns} S_{r,s}(n,k)(a^{\dagger})^{k}{a}^{k}  
\label{s3}
\end{equation}
or, alternatively,
\begin{equation}
[{x}^{r}(d/dx)^{s}]^{n}={x}^{n(r-s)}\sum_{k=s}^{ns} S_{r,s}(n,k){x}^{k}\,(d/dx)^{k}.
\label{s4}
\end{equation}
Eqs.(\ref{s3}) and ({\ref{s4}) introduce generalized Stirling numbers $S_{r,s}(n,k)$ which imply an extended definition of generalized Bell numbers:
\begin{equation}
 B_{r,s}(n) \equiv \sum_{k=s}^{ns} S_{r,s}(n,k).
\label{eb}
\end{equation}
Note that the Stirling numbers $S_{r,1}(n,k)$ have been previously  studied\cite{lan}. Also $B_{1,1}(n)=B(n)$ of Eq.(\ref{b1}). 

We have found a representation of the numbers $B_{r,s}(n)$ as an infinite series, which is a generalization of the Dobi\'{n}ski formula Eq.(\ref{b3}).  For $r=s$ one obtains:
\begin{equation}
B_{r,r}(n) = (1/e) \sum_{k=0}^{\infty}\frac{1}{k!}\left[\frac{(k+r)!}{k!}\right]^{n-1}, \; \; \; n=1,2 \ldots
\label{b4}
\end{equation}
with $B_{r,r}(0) = 1$ by convention.
For $r>s$ the corresponding formula is:
\begin{equation}
B_{r,s}(n) = [{(r-s)^{s(n-1)}}/e ] \sum_{k=0}^{\infty} \left[\prod_{j=1}^{s} \frac{\Gamma(n+\frac{k+j}{r-s})}{\Gamma(1+\frac{k+j}{r-s})}\right], \; \; B_{r,s}(0)=1
\label{b5}
\end{equation}

The formula Eq.(\ref{b3}) and its extensions Eqs.(\ref{b4}) and (\ref{b5}) share a common feature, namely, the fact that they give rise to a series of integers is by no means evident. 

Choosing various pairs $(r,s)$ gives alternative representations of many integer sequences.  Some examples are:
\begin{enumerate}
\item $(r>1,s=1)$ \\$B_{r,1}(n) = [{(r-1)^{n}}/e] \sum_{k=1}^{\infty} {\Gamma(n+\frac{k}{r-1})}/{[k!\, \Gamma(\frac{k}{r-1})]}$
\item The pair $(r+1,r)\; \;$ leads to a hypergeometric function $_{p}\!F_{q}$: \\
\begin{eqnarray}
B_{r+1,r}(n)&=&(1/e) [\prod_{j=1}^{r} \frac{(n-1+j)!}{j!}] \times \nonumber \\ & & \times _{r}\!F_{r}(n+1, n+2, \ldots, n+r;2,3,\ldots,r+1;1) 
\nonumber \end{eqnarray}
\item as does $(2r,r)$ \\
$B_{2r,r}(n)=[(rn)!/e\,r!] \, _{1}\!F_{1}(rn+1;r+1;1).$
\end{enumerate}

A still more general family of sequences arising from Eq.(\ref{b5}) has the form $(p,r=1,2,\ldots)$:
$$
 B_{pr+p,pr}(n)=(1/e)\left[\prod_{j=1}^{r} \frac{(p(n-1)+j)!}{(pj)!}\right] \times $$    
$$\times \,_{r}\!F_{r}(pn+1,\dots,pn+1+p(r-1);1+p,\ldots,1+p+p(r-1);1).
$$
For example, for $p=3,r=2$, this reduces to 
\begin{equation}
B_{9,6}(n)=(1/e)\frac{(3(n-1)+1)!(3(n-1)+2)!}{3!6!}\, _{2}F_{2}(3n+1,3n+4;4,7;1)
\label{b7}
\end{equation}
whose first four terms are 
$1,207775,566828686621,9011375448568566265.$ 

Knowledge of the generalized Stirling numbers in Eq.(\ref{s3})  solves the normal ordering problem for $[(a^{\dagger})^{r} a^{s}]^{n}\;$. We are able to give the appropriate generating functions for the sequences $B_{r,s}(n)$.  It then follows that, at least formally, we can furnish the generating functions for $S_{r,s}(n,k)$ as well\cite{pen}.  Additionally, it turns out that in certain circumstances  one may obtain explicit expressions for them.  We quote two such cases:
\begin{equation}
S_{r,r}(n,k)=\sum_{p=0}^{k-r} \frac{(-1)^{p}[\frac{(k-p)!}{(k-p-r)!}]^{n}}{(k-p)!p!}, \;\;\;\; (r\leq k \leq rn)
\end{equation}
and
\begin{equation}
S_{2,1}(n,k)=\frac{n!}{k!}\left(\matrix{n-1 \cr k-1\cr}\right), \; \; \;  (1 \leq k\leq n)
\end{equation}
which are the so-called unsigned Lah numbers\cite{com,lan}.
  
For those pairs $(r,s)$ for which we have an explicit expression for $S_{r,s}(n)$ we may generalize  Eq.(\ref{norm}) to obtain the normal ordered form of $e^{\lambda (a^{\dagger})^{r} a^{s}}$.  For example, the matrix element 
$\langle z |e^{\lambda (a^{\dagger})^{r} a}|z \rangle$   leads to the following normally ordered expression:
\begin{equation}
e^{\lambda (a^{\dagger})^{r} a} = {\mathcal N}(e^{\lambda (a^{\dagger})^{r} a}) = :\exp\{ [ (1-\lambda (a^{\dagger})^{r-1} (r-1))^{\frac{1}{r-1}} -1] a^{\dagger} a\}: 
\label{norm2}
\end{equation}
%%We are currently studying the feasibility of applying this approach to %%the general case $(r,s\neq 1)$.
We apply to Eq.(\ref{norm2}) the same method whereby we obtained the EGF of Eq.(\ref{EGF})  by taking the expectation of Eq.(\ref{norm}) in the coherent state $|z\rangle$, to get:
\begin{equation}
\langle z |e^{\lambda (a^{\dagger})^{r} a} |z\rangle = :\exp\{ [ (1-\lambda (z^{*})^{r-1} (r-1))^{\frac{1}{r-1}} -1] |z|^{2}\}: 
\label{expec}
\end{equation}
which evaluates at $z=1$ to give
\begin{equation}
\langle z |e^{\lambda (a^{\dagger})^{r} a} |z\rangle{_{z=1}} = :\exp\{  (1-\lambda  (r-1))^{\frac{1}{r-1}} -1 \}: \label{EGF2}
\end{equation}
which is precisely\cite{lan} the EGF for the numbers $B_{r,1}(n)$ .

For general $r>s$ the corresponding  $B_{r,s}(n)$ grow much more rapidly than $n!$ and thus may not be obtained via the usual form of EGF. One instead defines the EGF in terms of  $B_{r,s}(n)/(n!)^{t}$ where $t$ is an integer chosen to ensure that $\sum_{n=0}^{\infty}{B_{r,s}(n) }/{(n!)^{t+1}}$ has a non-zero radius of convergence.  As a result one obtains variants of Eq.(\ref{expec}) involving different hypergeometric functions\cite{pen}.
\section{Generalized Bell numbers as moments of positive measures}
The formulae Eq.(\ref{b4}) and (\ref{b5}) can be used to demonstrate what we consider to be the main thrust of our results, namely that $B_{r,s}(n)$ is the $n$-th moment of a positive probability measure.  For $r=s$ this takes the form of a sort of Dirac comb, while for $r>s$ gives a continuous distribution. 
Consider the case $r=s$ first.
We rewrite Eq.(\ref{b4}) as 
\begin{equation}
B_{r,r}(n) = (1/e) \sum_{k=0}^{\infty}\frac{1}{(k+r-1)!}[k(k+1)\ldots(k+r-1)]^{n} \; \; \; n=1,2\ldots
\label{b8}
\end{equation}
which immediately indicates that $B_{r,s}(n) $ is the $n$-th moment of 
\begin{equation}
W_{r,r}(x) = (1/e) \sum_{k=0}^{\infty}\frac{\delta(x-k(k+1)\ldots(k+r-1))}{(k+r-1)!}
\label{m6} \end{equation}
which is a  "rarefied" Dirac comb whose delta spikes are situated on the $x$-axis at $x=k(k+1)\ldots (k+r-1) \; \; \; k=0,1,\ldots$.

For $r>s$ it is necessary to excise the part of formula Eq.(\ref{b5}) which contains $n$, that is $\prod_{j=1}^{s} \Gamma(n+\frac{k+j}{r-s})$, for fixed $k$.  We reparametrize this with $n=\sigma -1$ and consider it as a Mellin transform; that is, we seek $\omega_{r,s}^{(k)}(x)$ such that when extended to complex $\sigma$
\begin{equation}
\int_{0}^{\infty}x^{\sigma-1} {\omega_{r,s}^{(k)}(x)} dx = \prod_{j=1}^{s} \Gamma(\sigma -1 +\frac{k+j}{r-s}).
\label{mel} \end{equation}
The positivity of ${\omega_{r,s}^{(k)}(x)}$ follows from the fact that it may be considered as resulting from an $s$-fold Mellin convolution of shifted Gamma-functions\cite{pen2,mar} which does preserve positivity.  We conclude that ${\omega_{r,s}^{(k)}(x)}$ is positive and so is $W_{r,s}(x)$, where 
\begin{equation}
\int_{0}^{\infty} x^{n} W_{r,s}(x) dx = B_{r,s}(n).
\label{sti}
\end{equation}

Use of the inverse Mellin transform yields many analytic forms of 
$W_{r,s}(x)$.  We exhibit some solutions of the Stieltjes moment problem Eq.(\ref{sti}):
\begin{enumerate}
\item For $r=2,3,\ldots$ 
\begin{eqnarray}W_{r,1}(x)&=&\frac{1}{e(r-1)}\left(\frac{x}{r-1}\right)^{\frac{2-r}{r-1}} \exp\left(-\frac{x}{r-1}\right) \times 
\nonumber\\
& &\times \sum_{k=0}^{\infty} \frac{1}{k!}\left(\frac{x}{r-1}\right)^{\frac{k}{r-1}}/{\Gamma\left(\frac{r+k}{r-1}\right)};
\nonumber
\end{eqnarray}
\item In particular 
\[
 W_{2,1}(x)=e^{-x} I_{1}(2\sqrt{x})/{(e \sqrt{x})},
\]
where $I_{1}(y)$ is the modified Bessel function of the first kind.  
\item Taking $r=3$ and $s=1$ we have $$
W_{3,1}(x)=\frac{1}{e\sqrt{8x}} e^{-\frac{x}{2}} \left[ \frac{2}{\sqrt{\pi}} \; \;_{0}\!F_{2}\left(\frac{1}{2},\frac{3}{2};\frac{x}{8}\right)
+\frac{x}{\sqrt{2}} \; _{0}\!F_{2}\left(\frac{3}{2},2;\frac{x}{8}\right)\right]
$$
\item while for $(r,s)=(5,2)$ we obtain  \begin{eqnarray}
 W_{5,2}(x)&=&\frac{2}{27 e}\frac{1}{\sqrt{x}} K_{\frac{1}{3}}(\frac{2\sqrt{x}}{3})\cdot u_{5,2}(x), \;{\rm where} \nonumber \\
 u_{5,2}(x&=&\frac{3}{32\pi}\left[ 24\sqrt{3} \;_{0}\!F_{4}\left(\frac{1}{3},\frac{2}{3},\frac{4}{3},\frac{5}{3};\frac{x}{243}\right)+ \right. \nonumber \\
    & & \mbox{} \left. +8\cdot 3^{\frac{5}{6}}x^{\frac{1}{3}}\; _{0}\!F_{4}\left(\frac{2}{3},\frac{4}{3},\frac{5}{3},\frac{6}{3};\frac{x}{243}\right)+ \right. \nonumber \\
    & & \left.+\mbox{} 3\cdot 3^{\frac{1}{6}}x^{\frac{2}{3}} \; _{0}\!F_{4}\left(\frac{4}{3},\frac{5}{3},\frac{6}{3},\frac{7}{3};\frac{x}{243}\right)\right]
\nonumber 
\end{eqnarray}
and $K_{\nu}(y)$ is the modified Bessel function of the second kind.
\item Finally
$$
W_{2r,r}(x)= \frac{1}{e \, r}x^{\frac{2-3r}{2r}}e^{-x^{\frac{1}{r}}}I_{r}(2x^{\frac{1}{2r}}).
$$
\end{enumerate}
Our proof of the existence and positivity of $W_{r,s}(x)$ implies the following \\{\em Theorem}: For all the sequences defined by Eqs. (\ref{b4}) and (\ref{b5}) the following inequalities are satisfied:
\begin{eqnarray}
\det\left[ B_{r,s}(i+j-2)_{1\leq i,j\leq n}\right] &>&0 \\
\det\left[ B_{r,s}(i+j-1)_{1\leq i,j\leq n}\right] &>&0 \; \; 
\mbox{ for all} \; \; n=1,2\ldots.
\end{eqnarray}
The above expresses the positivity of the two series of Hankel-Hadamard determinants spanned by the moment sequences.  This positivity is the necessary and sufficient condition for the existence of a positive measure $W_{r,s}(x)$ resulting from  the given sequence of moments\cite{akh}. 

For completeness we give here the two asymptotic expansions as $n \rightarrow \infty$ for the series $B_{2,1}(n)$ and  $B_{3,1}(n)$, obtained by the method of Hayman\cite{hay}.
$B_{2,1}(n) \equiv \langle z|[(a^{\dagger})^2 a]^{n}|z\rangle_{z=1}$  has the EGF
\begin{equation}
e^{\frac{x}{1-x}}=\sum_{n=0}^{\infty}{\frac{B_{2,1}(n)}{n!}}x^n
\end{equation}
and has the following asymptotics:
\begin{equation} 
B_{2,1}(n) \stackrel{n\rightarrow \infty}{\longrightarrow}\frac{1}{\sqrt{2e}}(n^{-\frac{1}{4}}+\frac{1}{12}n^{-\frac{3}{4}}+O(n^{-\frac{5}{4}}))\cdot n^{n} \exp(-n+2\sqrt{n}).
\end{equation}
Similarly, for $B_{3,1}(n) \equiv \langle z|[(a^{\dagger})^3 a]^{n}|z\rangle_{z=1}$
\begin{equation}
e^{\frac{1-\sqrt{1-2x}}{\sqrt{1-2x}}}=\sum_{n=0}^{\infty}{\frac{B_{3,1}(n)}{n!}}x^n
\end{equation}
\begin{equation} 
B_{3,1}(n) \stackrel{n\rightarrow \infty}{\longrightarrow}\frac{2^{\frac{1}{6}}}{\sqrt{3}e}(n^{-\frac{1}{3}}+2^{-\frac{3}{7}}n^{-\frac{2}{3}}+O(n^{-1}))\cdot (2n)^{n} \exp(-n+(3/2){(2n)}^{\frac{1}{3}}).
\end{equation}
The role played by these approximations is analogous to that of the Stirling approximation to $n$-factorial. 
The growth of other $B_{r,s}(n)$ with $n$ is in general even more rapid.  However, we do not have the asymptotic expansions for them at the present time.
\section{Postcript}
Our interest in integer combinatorial sequences (see also\cite{pen,pen2,pen3,six})  is a consequence of our parallel work on the construction of complete sets of coherent states (see\cite{kla2} and references therein).  Completeness - or, equivalently, the resolution of unity property - is an essential ingredient in such construction, and can be shown to be satisfied if one can furnish an appropriate positive measure $W(x)$.
Thus consider a state $|z\rangle$ constructed from an arbitrary discrete  set of states $|m\rangle$  which are both  orthonormal ($\langle m|m'\rangle=\delta_{m,m'}$) and complete:
\begin{equation}
|z\rangle = {N}^{-\frac{1}{2}}(|z|^{2}) \sum_{m=0}^{\infty} \frac{z^m}{\sqrt{\rho(m)}} |m\rangle,\; \; \rho(m)>0.
\label{res1}
\end{equation}
Resolution of unity is achieved via
 \begin{equation}
\int \!\!\int d^{2}z \,|z\rangle W(|z|^{2})\, \langle z| = I \equiv \sum_{m=0}^{\infty}|m\rangle \langle m|.
\label{res2}
\end{equation}
The conditions Eqs.(\ref{res1}) and (\ref{res2}) boil down to a relation closely resembling Eq.(\ref{sti}), namely
 \begin{equation}
\pi \int_{0}^{\infty} x^{n} \left[\frac{W(x)}{{N}(x)}\right] dx = \rho (n), \; \; \;x\equiv |z|^{2}\;\;\; n=0,1\ldots 
\label{res3}
\end{equation}
where it is assumed that the normalization of $N {(x)}$ defined by
 \begin{equation}
N(x)=\sum_{m=0}^{\infty} \frac{x^n}{\rho(n)}
\label{norm3}
\end{equation}
has a {\em finite} radius of convergence.   It is clear from Eq.(\ref{sti}) that when identifying $W_{r,s}(x)$ with $\pi W(x)/N(x)$ we can use the $B_{r,s}(n)$ as  $\rho(n)$  to construct  states of the form 
\begin{equation}
|z\rangle_{r,s} = {N}_{r,s}^{-\frac{1}{2}}(|z|^{2}) \sum_{n=0}^{\infty} \frac{z^n}{\sqrt{B_{r,s}(n)}} |n\rangle
\label{res4}
\end{equation}
where 
\begin{equation}
N_{r,s}(x)=\sum_{n=0}^{\infty} \frac{x^n}{B_{r,s}(n)}.
\label{norm4}
\end{equation}
As pointed out above, $B_{r,s}(n)$ grows much more quickly than $n!$ so the series Eq.(\ref{norm4}) is very rapidly convergent for all $r,s$. This ensures that the problem defined by Eq.(\ref{res3}) is well-defined for all $x>0$. 

The properties of the coherent states defined by Eq.(\ref{res4}) remain to be investigated but they automatically  satisfy the resolution of unity, 
due to  Eq.(\ref{sti}).

We are currently attempting to unravel the combinatorial meaning of some of the $B_{r,s}(n)$. It appears that this can be done systematically for  $B_{r,1}(n)$ at least, which will be the subject of a forthcoming publication\cite{men}. 

The integral representations of combinatorial numbers, used by us for the construction of coherent states, have triggered off applications for hyperdeterminants and Selberg integrals\cite{luq}. Although of evident utility, Stirling and Bell numbers have not yet found their way into most textbooks on mathematical physics, a notable exception being that of Aldrovandi\cite{ald}.
\section*{Acknowledgments}
We thank P. Blasiak, G. Duchamp, M. Mendez and J.Y. Thibon for enlightening discussions. 

\end{document}